# High-order harmonic generation in X-ray range from laser induced noble gas multivalent ions

JIXING GAO,[1,2] YINGHUI ZHENG,[1,2,4] JIAQI WU,[1,3] ZHIYUAN LOU,[1] FAN YANG,[1,2] JUNYU QIAN,[1,2] YUJIE PENG,[1,2] YUXIN LENG,[1,2] ZHINAN ZENG,[1,2,5] AND RUXIN LI,[1,3]

*1 State Key Laboratory of High Field Laser Physics, Shanghai Institute of Optics and Fine Mechanics, Chinese Academy of Sciences, Shanghai 201800, China*
*2 Center of Materials Science and Optoelectronics Engineering, University of Chinese Academy of Sciences, Beijing 100049, China*
*3 School of Physical Science and Technology, ShanghaiTech University, Shanghai 200031, China*
*4 e-mail: yhzheng@siom.ac.cn*
*5 e-mail: zhinan_zeng@mail.siom.ac.cn*

**Abstract:** Sub-femtosecond x-ray burst is powerful tool for probing and imaging electronic and concomitant atomic motion in attosecond physics. For years, x-ray source (above 2 keV) had mainly been obtained from X-ray free electron laser (XFEL) or synchrotron radiation, which are high energy consumption, high cost and huge volume. Here we propose a low-cost and small-size method to generate X-ray source. We experimentally obtained high photon energy spectrum (~ 5.2 keV) through both atom and multiple valence state ions using a near-infrared 1.45 μm driving laser interacting with krypton gas, according to our knowledge, which is the highest photon energy generated through high-order harmonic generation up to now. In our scheme, multi-keV photon energy can be achieved with a relaxed requirement on experimental conditions, and make time-resolved studies more accessible to many laboratories that are capable of producing high energy photon extending to hard x-ray region. Furthermore, our scheme minimizes the influence of X-ray fluorescence process on detection, and can also be utilized to study the quantum-optical nature of high-order harmonic generation.

## 1. INTRODUCTION

Real-time observation of intra-atomic electron dynamics requires the ultra-short time spans from femtoseconds to attosecond, which is equivalent to energy scale spans from 0.1-10 keV. Extreme ultraviolet (XUV) and x-ray attosecond bursts are considered to serve as the pump and the probe events in ultrafast dynamics [1-8]. The high-order harmonic generation (HHG) is used as mainly tabletop coherent light source in attosecond technology but the most applications are limited in the XUV region ( < 150 eV) [9]. X-rays can acquire the internal state of atoms and molecules motions (the elements of materials, spin state, charges in absorption or coupled motions, etc.) [10], which need the photon energy extend to above ~ 500 eV and even to keV region [11].

The physical mechanism of HHG is well explained by the semi-classical three-step model [12], which is once the electrons are excited from gas phase atom by tunneling ionization, then they will be accelerated in the laser field, finally come back to recombine with parent ions with gained energy, and the result of electron recolliding with ion mainly towards three channels: high-order harmonic electromagnetic radiation, elastic scattering and inelastic scattering. The returning electron with enough high kinetic energy will initiate excitation or secondary ionization of the atom or molecule. According to the model, the highest photon energy of HHG cutoff region is given by the equation: $h\nu_{cutoff} = I_p + 3.17 U_p$, where $I_p$ is the ionization energy of selected atomic gas, and $U_p \propto I_L \lambda_L^2$ is the ponderomotive potential of liberated electrons gained during the acceleration process in the laser field with the laser intensity $I_L$ and wavelength $\lambda_L$. Hence, there are two main directions to gain higher photon energy: using a laser field with longer wavelengthor with higher intensity.

On the one hand, since the highest generated photon energy is proportional to the square of driven laser wavelength ($\propto \lambda_L^2$) [12], many schemes to generate high-order harmonics driven by mid-infrared (MIR) laser have been proposed. Chen et al produced ~540 eV HHG through phase-matching (PM) by focusing 40 fs / 2 μm laser into He gas-filled hollow waveguide with high gas pressure [11]. Teichmann et al. generated high harmonic radiation in soft X-ray region about 550 eV by carrier envelope phase stable 12 fs / 1.85 μm laser pulses with 5×10$^{14}$ W/cm$^2$ interacting with 12.5 bar He gas filled cell [13]. Popmintchev et al. achieved more than 1.6 keV HHG bursts driven by focusing six-cycle 3.9 μm MIR pulses at an intensity of $I_L$ = 3.3×10$^{14}$ W/cm$^2$ into a hollow waveguide filled with helium gas [14]. Zenghu Chang theoretically proposed to produce ~4 keV photon energy using 8 μm laser pulses interacted with He gas which can be further enhanced more than two orders of magnitude by additional 0.8 μm laser [15]. But there is an inevitable disadvantage in those schemes that is the conversion efficiency of HHG decreases rapidly to the -5.5 power of the driving laser wavelength ($\propto \lambda_L^{-5.5}$) [16,17], and thus extra-high gas pressure is usually required to improve the conversion efficiency of high photon energy region which increases the complexity of the experiment. Hence, considering the balance between cutoff photon energy and conversion efficiency, compared with the long wavelength of MIR laser, near-infrared (NIR) driving laser is a more appropriate choice. On the other hand, the ponderomotive potential can be directly improved by increasing the laser intensity, and PM is optimized with the increasing laser intensity under ionization saturation intensity [18]. When the laser intensity beyond the ionization saturation intensity, the contribution of gas plasma and clusters will become more important which produce x-ray fluorescence. Dobosz et al. obtained 2.5 keV x-ray spectrum by using 0.79 μm infrared laser with $5×10^{17}$ W/cm$^2$ interacting with Kr gas [19]. Mcpherson, et al. generated keV x-ray spectrum from Kr Clusters with 0.248 μm driving laser at the peak intensity of $8×10^{18}$ W/cm$^2$ [20]. However, these have not been proved to be high-order harmonic spectra

Recently, mutivalent ions represent attractive medium for efficient XUV and soft X-ray HHG. S. G. Preston, et al. [21] obtained 37th harmonic by 248.6 nm, 380 fs KrF laser with intensity $4×10^{17}$ W/cm$^2$, they using the single atom response theoretical modeling that identifies the ion species He$^{1+}$, Ne$^{1+}$, and Ne$^{2+}$ as the medium of the harmonics. But they produced HHG emission at very low (few torr) gas pressures, which means X-rays from ions in a non phase matching (NPM) regime [16, 22]. However, in 2015, Popmintchev et al [23]. reported to achieve 280 eV HHG by using 35 fs / 267 nm lasers with > $6×10^{15}$ W/cm$^2$ peak intensity with Ar (about 400 torr) multiply ionized plasma in waveguides and they identified both atoms and multiply ions can enable effective PM, that the mutivalent ions have the great possibility to produce high energy photons (multi- keV) through HHG process. These aricles show that gas ions can produced HHG emission by NPM regime and PM regime, more importantly, they prove the feasibility of mutivalance ions HHG mechanism. But the harmonics with higher photon energy (muti-keV) beyond the saturated laser intensity have not been observed and proved experimentally up to now.

In this work, we report the experimental evidence of HHG with cutoff energy up to 5.2 keV for the first time to our knowledge by focusing 1.45 μm NIR laser into a Kr gas at laser intensity approaching $2.5×10^{16}$ W/cm$^2$. In our experiment the wavelength of the driving laser is relatively short compared with the MIR and the gas pressure is also relatively low, only 700 torr, which reduces the requirements of the experimental. The driving laser intensity is high enough to excite multivalent ions to participate in the harmonic generation process. The theoretical simulation results show that atoms and ions as high as eight valence participate in the harmonic process. Meanwhile the incidence energy is high enough to achieve required intensity to excite gas ions taking part in the generation of high energy x-ray photons [21, 23-30]. In general, the high photon energy spectrum (above keV) through HHG is weak and also accompanied with continuous X-ray spectrum (e.g., bremsstrahlung) [31, 32], characteristic X-ray spectrum [33], X-ray fluorescence spectrum [34] and other mechanics [20, 35] (e.g. inverse bremsstrahlung (IB) [36], above threshold ionization (ATI) [37]), hence it is difficult to separate the HHG spectrum and directly measure the coherence by flat-field grating spectrometer. Our solution is to minimize the influence of X-ray fluorescence process, then employ the single photon detector to collect HHG signal. In the future, our scheme can also be utilized to research experimentally the quantum-optical nature of high-order harmonic generation [38].

## 2. EXPERIMENTAL SETUP

As shown in Fig. 1(a), in our experiments we use a NIR laser system based on OPCPA that delivere 7.5 mJ , 60 fs ,1.45 μm pulses with a repetition of 20 Hz [39]. This beam is focused into a vacuum chamber using a CaF2 lens with 100 mm focal length with a focal diameter of ~12.5 μm. Kr gas are injected into the vacuum chamber with a pulsed nozzle (Series

9 Parker Pulse Valves, orifice diameter of 300-500 μm) exit mounted on a three-dimensional motorized translation stage. The distance between nozzle and laser beam is 100-1000 μm and make sure the laser does not hit the jet. The additional diaphragm (a diameter of ~3mm) is inserted between the gas jet and the detector to block the undesired signal (partial fluorescence spectrum and NIR laser). The generated x-ray spectra are recorded by an x-ray detector filtered with 8-μm-thick aluminum(Al), which is mounted ahead of the 8-μm-thick beryllium(Be) window of the detector, so signals below 0.75 keV are filtered out and the count rate is maintained below 1 counts/s This detector is a single photon detector which can not only increase the detective efficiency but also be convenient to change the detective angle that is achieved by placing both the diaphragm and the detector on a rotation stage. A pair of magnets are placed in front of the Al filter to deflect free electrons, which are accelerated to high energy, and then prevented them to strike a metallic target. The metal elements in the light path are all wrapped in black paper to prevent electrons from bombarding metal. Furthermore, the whole vacuum chamber and the external optical table are linked to earth in order to shield electromagnetic interference from outside electrical facilities such as vacuum pumps and electrical controllers.

The x-ray detector is calibrated using the fluorescence peak of iron in Shanghai Synchrotron Radiation Facility (SSRF) [40, 41]. The energy resolution at the 5.9 keV peak is 125 eV (full-width at half maximum, FWHM). The detector works in a photon-counting mode and both the photon number and photon energy can be recorded.

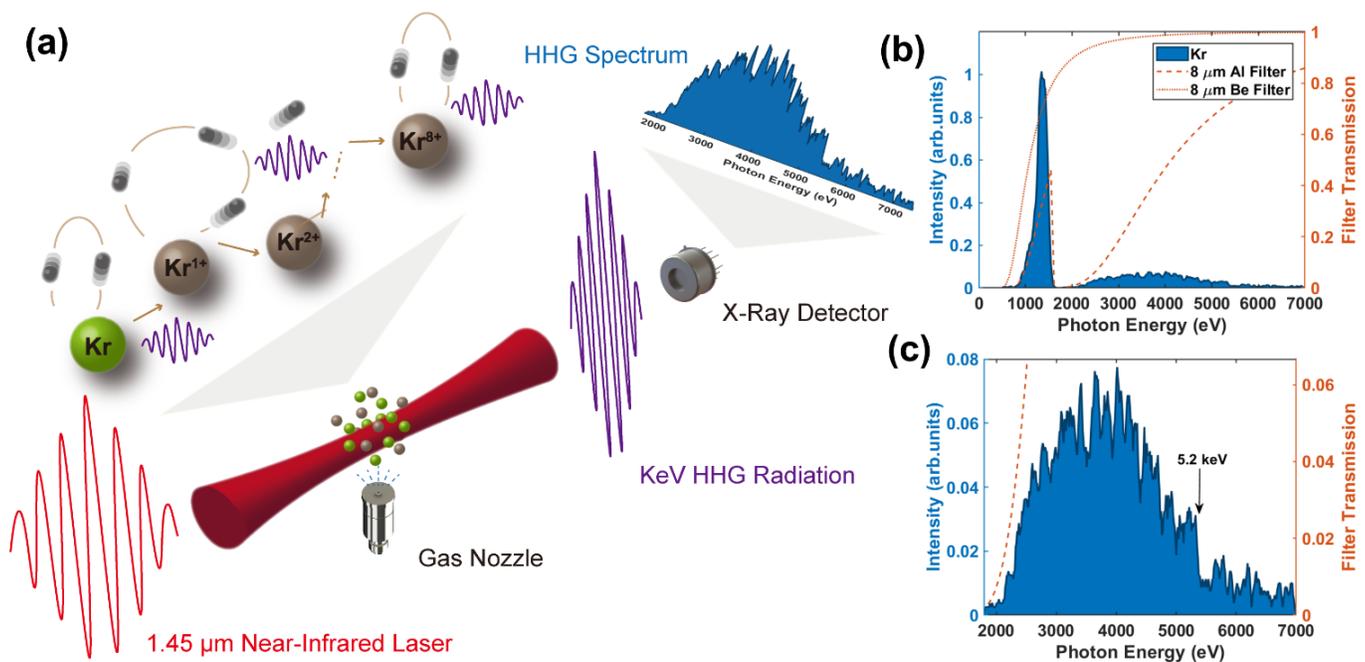

Fig. 1. (a)The schematic of HHG produced through NIR laser interacted with Kr atom and multiple valence ions. The highest valence state of Kr that make a contribution to the HHG spectrum is Kr8+. (b)The HHG spectrum from Kr gas interacted with linear polarized incidence laser with 7.5 mJ energy. The detector has a minimum energy cutoff 0.75 keV due to the Be window. (c) Enlargement of (b) range from 2 keV to 7 keV. The red dotted and dash curves are the transmission curves of 8-μm-thick Al filter and 8-μm-thick Be window, respectively.

## 3. RESULTS AND DISCUSSION

**A. HHG with 5.2 keV photon energy**

As can be seen in Fig. 1(b) the whole experimental spectrum ranges from 0.75 -5.2 keV, of which less than 0.75 keV is filtered by the Be window in front of the detector, and 1.55 keV is the K-edge of Al. The dotted line in the Fig. 1(b) is the transmittance curve of 8 μm-thick Be window, and the dotted line is the transmittance curve of 8 μm-thick Al film. Fig.1(c) shows the amplification of the 2 keV-7 keV part of the spectrum in Fig. 1(b). It can be seen that there is an obvious cutoff at 5.2 keV. In order to avoid the two-photon or even three-photon effect of the detector, we use an 8 μm-thick Al film in front of the detector window to filter out the high counting low-energy photons and ensure that the photon counting rate is lower than 0.05 photons/s/pulse. The other reason that we choice 8-μm-thick Al as the filter is the absorption edge of Al (K edge) has the minimum of transmitted intensity of Kr characteristic X-ray radiation, which will reduce the influence of L series lines emission. Through such processing, the experimental error of high-order harmonic detection is minimized as much as possible. Here, the experimental laser intensity is $I_{max}$ = 2.5×10$^{16}$ W/cm$^2$ that is high enough to excite M shell electrons of Kr. High intensity laser field will excite multiple ionization processes that play a critical role in the generation of harmonics with high photon energy. Although He gas has a higher ionization potential, it has exceptionally small effective nonlinearity [42], so extremely high gas pressure is required in order to obtain keV photon energy, and increases the difficulty of experiment. For example, Popmintchev et al. produced 1.6 keV photon energy spectrum with 35 atm He gas. However, we generate 5.2 keV photon energy with only ~700 torr gas pressure.

**B. Theoretical analysis**

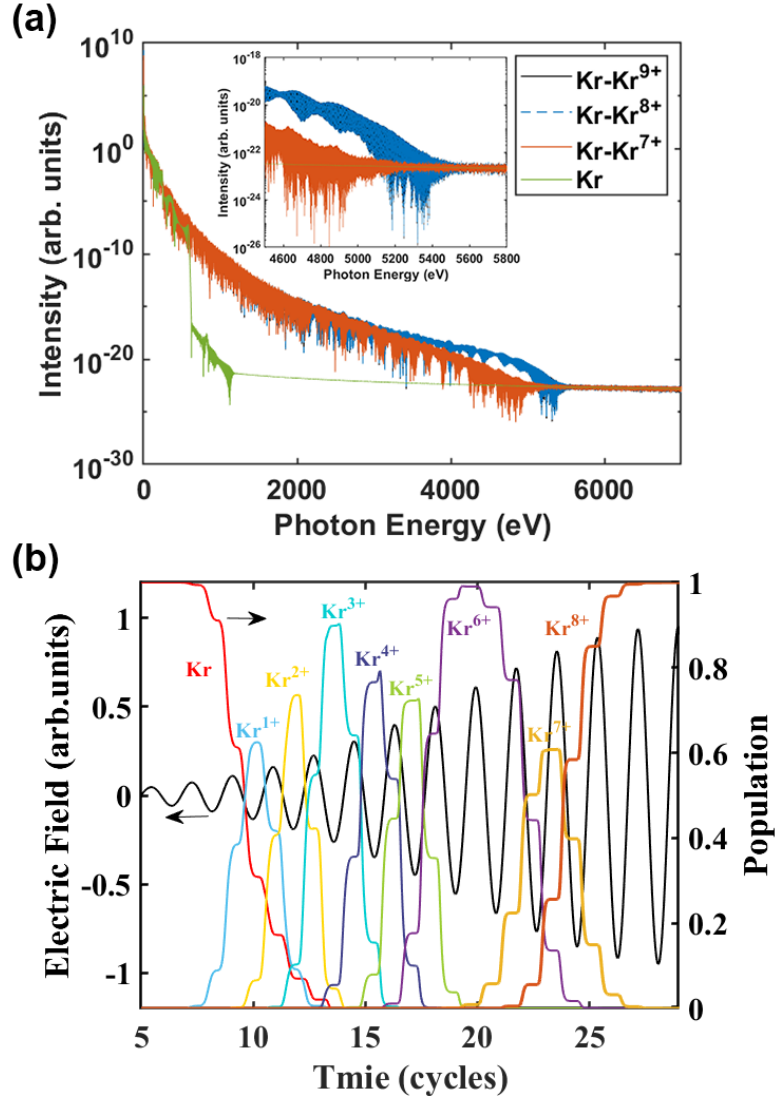

Fig. 2 (a) HHG spectra of the Kr atoms (green line) and multiple Kr ions (red solid line: from Kr to Kr[7+], blue dotted line: from Kr to Kr[8+], black solid line: from Kr to Kr[9+]) from 1D-TDSE calculation. The inner panel is the enlargement of HHG spectrum around cutoff energy. (b)Evolution of the different ion populations with the laser field (black line). The populations of different ions are plotted separately. The intensity peak of laser is at the 29th cycle. In simulation, the duration, wavelength and intensity of laser field are 60 fs (FWHM) 1.45 μm and $2.5\times10^{16}$ W/cm$^2$, respectively.

---

In order to understand the generation of such high energy photon, we numerically study multivalent Kr ions interacted with laser field through solved 1D time-dependent Schrödinger equation (TDSE). Here we treat the atom and ions in a Single-electron approximation (SAE) [29, 43]. Then the 1D-TDSE is written as (in a.u. $e=\hbar=m_e=1$)

$$i\frac{\partial}{\partial t}\psi(x,t) = \left[-\frac{1}{2}\frac{\partial^2}{\partial x^2} + V(x) - xE(t)\right]\psi(x,t), \tag{1}$$

Where $V(x) = -\frac{z}{\sqrt{a^2+x^2}}$ is Soft-core Coulomb potential, with z represent the nuclear charge and α is the softening parameters. In calculations, we need to change the value of a to make the ground state energy of particle equal to the ionization energy (through solve the stationary Schrodinger equation) for atom and different valance ions (i.e., different z). $E(t) = E_0 f(t)\sin(\omega_0 t)$ is the laser electric field, with $E_0$ is the peak amplitude of the laser field, we employ an Gaussian envelope function $f(t) = e^{-4ln2[(t-t_0)/T]^2}$, where T is the pulse duration.

Fig. 2(a) is the high-order harmonic spectra simulated from the Kr atoms (green line) and multiple Kr ions (red solid line: from Kr to $Kr^{7+}$, blue dotted line: from Kr to $Kr^{8+}$, black solid line: from Kr to $Kr^{9+}$). There is excellent agreement between the theoretically calculation and the experimental results. In order to simulate closer to the real situation, we consider the next two conditions (as shown in Fig. 1(a)). First, we consider that only Kr atoms are involved in the harmonic generation process. As shown the green solid curve in the Fig. 2, the cutoff of the harmonic is 0.6 keV. Then, it is considered that ions from $Kr^{1+}$ to $Kr^{16+}$ are added to the harmonic generation process in turn. We find that when the simulation parameters such as driving laser intensity and pressure are the same as those in experiment, the maximum cutoff energy is ~5.2 keV. At this time, atoms and ions from $Kr^{1+}$ to $Kr^{8+}$ participate in harmonic generation and the $Kr^{8+}$ plays an important role in photon energy extending to ~5.2 keV, while $Kr^{9+}$ and ions with higher valence state almost have no contribution to the harmonic spectrum. This result is in good agreement with the experiment, so we infer that in our experiment, in addition to atoms, a total of 8 valence ions contribute to harmonic generation that means the N-shell electron (the 4s and 4p shell) of Kr takes part in HHG process.

In order to further explain multivalent ions HHG process, we numerically solve the ionization of atom and ions from TDSE. Fig. 2(b) shows the populations of Kr atom and multivalent ions (Kr to $Kr^{m+}$, with m=1, 2, ..., 16) as a function of time along with the laser field, indicated by red, blue, yellow, cyan, slateblue, green, purple, orange, maroon solid curves respectively, where the black curve indicates the laser field. From Fig. 2(b) we can see that with the time evolution of the laser field, Kr atoms and $Kr^{1+}$ to $Kr^{8+}$ ions appear at different times of the laser field, and they all have a population of more than 0.6. Among them, Kr atoms and $Kr^{1+}$ to $K^{7+}$ ions appear at the front of the laser pulse. Corresponding to the spectra of Fig. 2(a) which are obtained by multiplying the dipole moment by this population evolving with time and performing Fourier transform, Kr atoms mainly produce low photon energy spectrum below 0.6 keV (green line), and other low valence ions mainly contribute to the platform of spectrum (red solid line). $Kr^{8+}$ ion appears near the peak of the laser field, so in Fig. 2(a) it mainly contributes to the cut-off region of HHG (blue dotted line). The ionization energy of the electron in the 3d shell corresponding to the 9-valent ion is very high and it is difficult to ionize. We calculate that the population of the $Kr^{9+}$ is 0.0001, so the ions of $Kr^{9+}$ and higher valence state have no contribution to HHG, as shown in Fig. 2(a), the spectrum of Kr-$Kr^{9+}$ (black solid line) and Kr-$Kr^{8+}$ (blue dotted line) is completely overlapped.

## C. HHG mechanism

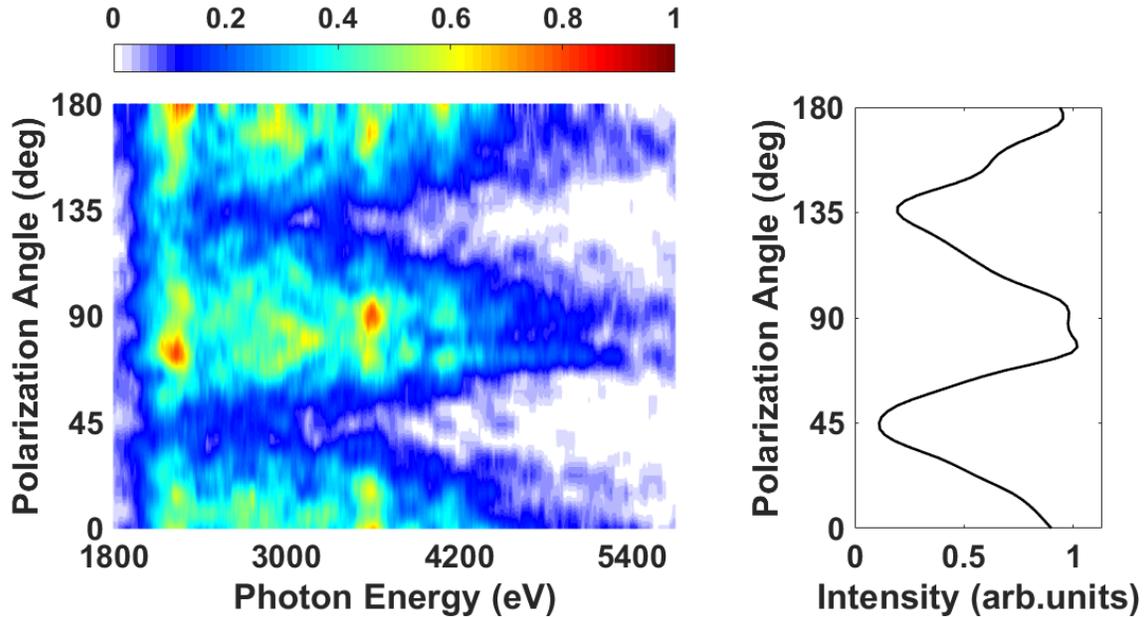

Fig. 3. The dependence of HHG spectra with polarization angle. The pseudo-color map of HHG signals with the polarization angles rotated from 0° to 180° with a λ/4 laser plate. The 0°, 90° and 180° represents linear polarization laser, 45° and 135° represents circular polarization laser, respectively. The experimental conditions remain the same as those in Fig.1.

Next, we need to verify that the spectrum with multi-keV photon energy is truly from HHG mechanism instead of continuous and characteristic X-ray spectrum, X-ray fluorescence spectrum and other mechanics, e.g., inverse bremsstrahlung (IB), and above threshold ionization (ATI). First, one of the characteristics of HHG signals is extremely sensitive to the ellipticity of the incident laser. The harmonic signal is the strongest when the laser is linearly polarized. With the increase of ellipticity, the harmonic intensity decreases exponentially [44]. Therefore, we test the ellipticity dependence of the HHG spectrum. As shown in Fig. 3, the 0°, 90° and 180° positions of the λ/4 wave plate correspond to the linear polarization, while 45° and 135° stand for the circular polarization, respectively. We can see the signal varies periodically as the ellipticity changes, and the linearly polarized laser produces a much stronger signal than that driven by the circularly polarized laser. It is in well accordance with the nature of gas HHG process [41, 44-47].

Second, the HHG signal has strong spatial directivity [45, 48], while x-ray fluorescence emission which appears to be isotropic [19]. Thus, with horizontally rotating the angle of detector, we can obtain the angular distribution of the x-ray emission ranging from 2 keV to 5.5 keV as shown in Fig. 4. It can be seen that the signal is approximately symmetrical angular distribution centered on the driving laser propagation direction. According to the Gaussian fitting curve, the full-width at half maximum of the angular distribution is 12.8 degrees.

Comparing to the divergence of the conventional PM gas HHG, our result has a larger spatial angular distribution that possibly caused by the following reasons. According to the previously reported theoretical and experimental results [49-51], the dipole moment dominates gas HHG process with optimized PM conditions, which means HHG signal has a strongly directional emission along the driving laser propagation direction. However, when the laser intensity is increased beyond the saturation intensity of ionization and even reaches the relativistic intensity, the dipole approximation will be broken and the multipolar effect (e.g., quadrupolar) will play more important role in HHG. The nondipole effect will provide sufficiently conversion efficiency of HHG at multi- keV photon energies). Some theoretical papers have already confirmed that when the laser intensity is high enough ($> 10^{15}\ W/cm^2$) and the photon energy reaches keV region, multipolar effect will become essential [38, 52-55]. Recently, Alexey Gorlach et. al. proposed that a full-quantum theory analysis method to explain gas HHG process [38]. They predicted that with the increase of laser intensity, the contribution of multipole radiation to harmonics increases rapidly compared with dipole radiation. Especially, for the harmonics with photon energy above keV, the quadrupolar radiation becomes comparable to the dipolar radiation and even the multipolar effect need be further considered. Furthermore, nondipole effect has a different spatial angular distribution from conventional PM dipole effect. Therefore, the nondipole effect not only can generate harmonics with multi-keV photon energy, but also has a larger spatial angular distribution. In this paper, we preliminarily analyze that the HHG spectrum is related to the non-dipole effect with high driving laser intensity. This part is the content of the following research and still needs to be further discussed.

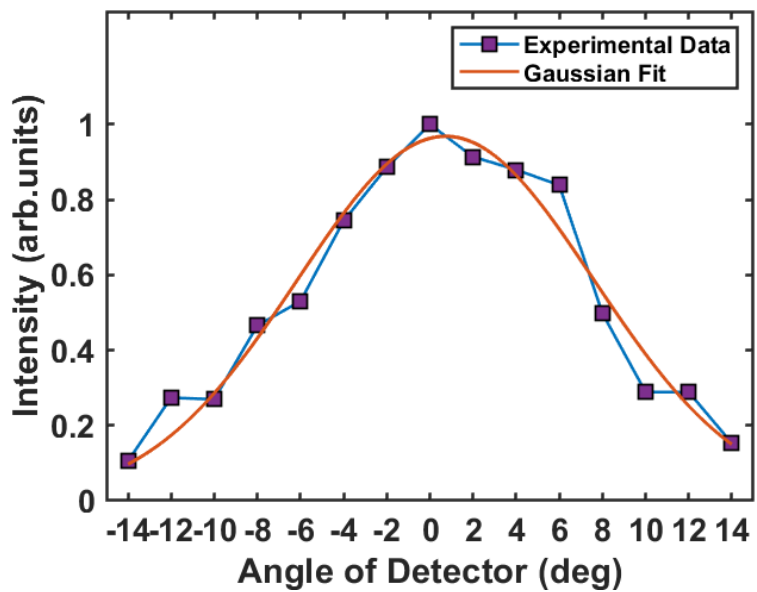

Fig. 4. The angle distribution of the x-ray emission ranging from 2 keV to 5.5 keV. 0° represents the direction in which the laser propagation.

## 4. SUMMARY

In conclusion, we propose a low-cost and small-size method to generate X-ray source. High photon energy spectrum (~ 5.2keV) is obtained through both atom and multiple valence state ions, according to our knowledge, which is the highest photon energy generated through HHG up to now. Both the distinct spatial angular distribution and the ellipticity dependence of signal are strong evidence of HHG mechanism. When driven laser intensity exceeds saturation intensity of ionization, and the dipole approximation will be broken which means dipole radiation may not be the only contribution to HHG, and the multipolar radiation plays an important role under this circumstance. In our scheme, the HHG x-ray source with multi-keV photon energy can be achieved with a relaxed requirement on experimental conditions, and be more accessible to many laboratories. Furthermore, our scheme minimizes the influence of X-ray fluorescence on detection, and can also be utilized to study the quantum-optical nature of high-order harmonic generation.

**Funding.** This research was funded by the National Natural Science Foundation of China (grant No. 61690223, No. 11127901, No. 91950203, and No. 11874374), the Strategic Priority Research Program of the Chinese Academy of Sciences (grant No. XDB16), and the Chinese Academy Sciences (CAS) Youth Innovation Promotion Association.

**Disclosures.** The authors declare no conflicts of interest.

**Data availability.** Data underlying the results presented in this paper are not publicly available at this time but may be obtained from the authors upon reasonable request.